\documentclass[bibnotes,prl,amsmath,twocolumn,floatfix,citeautoscript,
               preprintnumbers,amssymb,amsmath,showpacs,
               letterpaper]{revtex4}
\pdfoutput=1
\usepackage{graphicx} % to include images
\usepackage{microtype}
\begin{document}
% -----------------------------------------------------------------------------
\title{Electron-lattice instabilities 
       suppress cuprate-like 
       electronic structures in SrFeO$_3$/SrTiO$_3$ 
       superlattices}
  \author{James M.\ Rondinelli}
     \email[Address correspondence to: ]{rondo@mrl.ucsb.edu}
  \affiliation{Materials Department, University of California, Santa Barbara, 
	       CA, 93106-5050, USA}
  \author{Nicola A.\ Spaldin}
  \affiliation{Materials Department, University of California, Santa Barbara, 
	       CA, 93106-5050, USA}
\date{\today}
% -----------------------------------------------------------------------------

% -----------------------------------------------------------------------------
\begin{abstract}
Using {\it ab initio} density functional theory we explore the behavior of 
thin layers of metallic $d^4$ SrFeO$_3$ confined between the $d^0$ dielectric 
SrTiO$_3$ in a superlattice geometry.
We find the presence of insulating SrTiO$_3$ spacer layers strongly 
affects the electronic properties of SrFeO$_3$:
For single SrFeO$_3$ layers constrained to their bulk cubic structure, 
the Fermi surface is two-dimensional, nested and resembles the  
hole-doped superconducting cuprates.
A Jahn-Teller instability couples to an octahedral tilt mode, however, 
to remove this degeneracy resulting in insulating superlattices.
\end{abstract}
% -----------------------------------------------------------------------------

% -----------------------------------------------------------------------------
% -----------------------------------------------------------------------------
\maketitle
% -----------------------------------------------------------------------------

\begin{figure}
\includegraphics[width=0.44\textwidth]{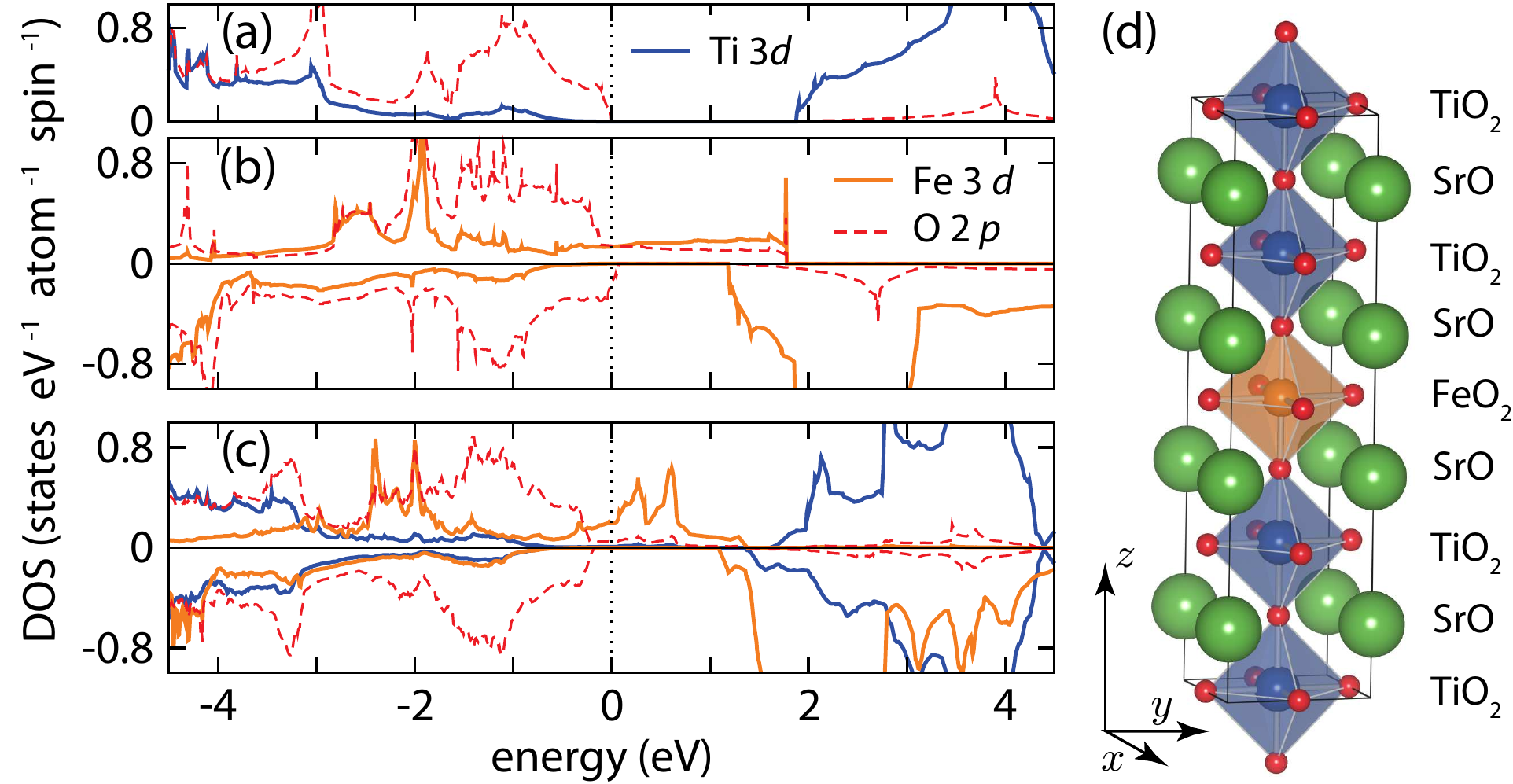}
\caption{\label{fig:3sto-1sfo-dos} (Color) %
The DOS for 
(a) cubic SrTiO$_3$  ($a=3.86$~\AA), (b) cubic FM SrFeO$_3$  ($a=3.79$~\AA), 
and (c) the high symmetry $P4/mmm$ (SrTiO$_3$)$_3$/(SrFeO$_3$)$_1$ 
superlattice shown in (d).}
\end{figure}

Progress in the layer-by-layer growth of transition metal oxide thin 
films \cite{Hwang:2006,Reiner/Walker/Ahn:2009} motivated the intriguing 
recent suggestion that oxide heterostructures engineered to have band 
structures close to those of the high-$T_c$ cuprates could yield new 
superconductors \cite{Chaloupka/Khaliullin:2008}.
Strained LaNiO$_3$ layers separated by inert spacers such as 
LaAlO$_3$ or LaGaO$_3$ were proposed as a promising trial system, 
with strain lifting the degeneracy of the single Ni$^{3+}$ $e_g$ electron 
and layering providing quasi two-dimensionality.
Indeed, first-principles electronic structure 
calculations \cite{Hansmann/Held_et_al:2009} on strained Ni-based oxide 
superlattices have found that, with careful choice of strain and 
local-interface chemistry \cite{Anderson/APS:2009}, 2D Fermi surfaces
resembling those of the hole-doped 
cuprates \cite{Pavarini/Andersen_et_al:2001} can be obtained.
Superconductivity in such superlattices remains to be observed experimentally, 
however, possibly because competing instabilities such as charge- or 
orbital-ordering are enhanced by the reduced dimensionality. %\cite{Keimer:2008}
Although high-$T_c$ superconductivity tends to occur in 
materials with large electronic fluctuations and in close proximity to 
electronic, structural or magnetic 
phase transitions \cite{Morosan/Cava_et_al:2006,Bianchi/Fisk_et_al:2008,Nagaosa:1997,delaCruz/Pengcheng_et_al:2008}, the detailed role of such instabilities remains 
unclear \cite{Emery/Kivelson:1995,Allen:1982}.
With these factors in mind, we examine how structural confinement and 
lattice instabilities modify the electronic structure of SrFeO$_3$ 
in SrTiO$_3$/SrFeO$_3$ superlattices.
Our motivation for this choice of system is three-fold: First, like 
Ni$^{3+}$, Fe$^{4+}$ has a single degenerate $e_g$ electron in an
octahedral environment that dictates its low energy physics. 
The restriction on Fe-based compounds---traditionally dismissed from 
consideration for superconductivity because of their robust magnetism---has 
been lifted due to the recent discovery of superconducting  
Fe pnictides \cite{Hosono/FeSC:2008}.
Second, bulk SrFeO$_3$ is metallic with $p$-type conductivity (like the doped 
cuprates) and is proximal to multiple instabilities: It manifests a 
long-wavelength spin density wave, but neither Jahn-Teller distorts nor 
charge orders, even though both possibilities are suggested by its chemistry.
Finally, unlike the structurally inert LaAlO$_3$ in the nickelate 
superlattices, SrTiO$_3$ is a highly polarizable dielectric which can couple 
to electronic or structural distortions \cite{Okamoto/Millis/Spaldin:2006} 
in the SrFeO$_3$ layer.
Using first-principles density functional theory (DFT) within the local-spin 
density approximation (LSDA) plus Hubbard $U$ method, we calculate the 
structure and electronic properties of (SrTiO$_3$)$_n$/(SrFeO$_3$)$_m$ 
superlattices. We focus on ($i$) the evolution of the 2D band structure 
with thickness of the dielectric and metal ($n,m=1$ or 3), 
and ($ii$) how competing structural and electronic instabilities manifest 
in the superlattices. 
We find that 2D confinement from the superlattice periodicity along 
the growth ($z$) direction yields low-energy physics that are primarily derived 
from hybridized Fe $d_{x^2-y^2}$ and O 2$p$ orbitals.
The idealized high-symmetry superlattice structure  
has a strongly nested 2D Fermi surface that is similar to that of  
the parent superconductor La$_2$CuO$_4$ \cite{Cohen/Pickett/Krakauer:1988}. 
Strong electron-lattice instabilities, enhanced by 2D confinement, 
transform the metastable metallic structure into a lower 
symmetry {\it insulating} superlattice.
Understanding the competing nature of these often neglected lattice distortions 
is crucial for engineering electronic structures of oxide 
superlattices and interfaces. 
Our LSDA$+U$ DFT calculations are performed using the 
Vienna {\it ab initio} Simulation Package 
({\sc vasp}) \cite{Kresse/Furthmueller_PRB:1996,Kresse/Joubert:1999}.
We follow the Dudarev approach \cite{Dudarev_et_al:1998} and include 
an effective Hubbard term $U_{\rm eff}=U-J$ of 6~eV to treat the  
Fe 3$d$ orbitals \footnote{%FOOTNOTE ON METHODS
The core and valence electrons are treated with the projector augmented 
wave method \cite{Bloechl:1994} with the following valence electron 
configurations: 
$3s^23p^64s^2$ (Sr), $3p^63d^74s^1$ (Fe), $3s^23p^63d^24s^2$ (Ti) and $2s^22p^4$ (O). 
The Brillouin zone integrations are performed with a Gaussian smearing of 
0.05~eV over a $9\times9\times5$ Monkhorst-Pack $k$-point 
mesh \cite{Monkhorst_Pack} centered at $\Gamma$, and a 450 eV plane-wave cutoff. 
For structural relaxations, we relax the ions until the Hellmann-Feynman forces 
are less than 1 meV \AA$^{-1}$.
}.  %%%-FOOTNOTE ON METHODS
This method and value of $U_{\rm eff}$ gave good results in earlier 
first-principles calculations for bulk SrFeO$_3$ and related iron oxide  
compounds \cite{Saha-Dasgupta/Popovic/Satpathy:2005,Shein/Ivanovskii_et_al:2005}.
%,Rondinelli/Spaldin:2009}
%
We construct the superlattices by stacking %$n=1\ldots4$ 
five-atom perovskite %SrTiO$_3$ 
units along the $z$-direction % followed by $m$ unit cells of SrFeO$_3$ 
[see Fig.\ \ref{fig:3sto-1sfo-dos}(d)], and 
constrain the in-plane lattice parameter ($xy$-plane) to that of  
cubic LDA SrTiO$_3$ ($a=3.86$~\AA) to simulate growth on a SrTiO$_3$ substrate. 
While the lattice mismatch between SrFeO$_3$ and SrTiO$_3$ is small 
(the theoretical mismatch is 1.8\%, with ferromagnetic LSDA$+U$ SrFeO$_3$ 
having $a=3.79$~\AA), we show below that it is sufficient to impose an 
epitaxial crystal field ($\Delta_{\rm ECF}$) that partially lifts the Fe 
$e_g$ orbital degeneracy. 
%
%Periodic boundary conditions are used to create a repeating superlattice.
%
We then relax the length of the $c$-axis and the internal degrees of 
freedom along the $z$-direction. 
The in-plane periodicity imposed by this choice of reference 
structure does not permit cell-doubling structural distortions
 such as octahedral rotations or orbital ordering:
This highly-symmetric superlattice (space group $P4/mmm$) resembles the 
structural constraints often considered to be imposed in heteroepitaxial thin films.
We relax this restriction later to study electron-driven lattice instabilities.
Ferromagnetic (FM) order is imposed on the Fe sites in all 
calculations, and is theoretically \cite{Shein/Ivanovskii_et_al:2005} %,Rondinelli/Spaldin:2009}  
found to be the lowest energy collinear ordering for bulk SrFeO$_3$.
%\footnote{The experimentally observed antiferromagnetism results from a long 
%wavelength screw structure in which the angle between neighboring spins is 
%$\sim$40$^\circ$.\cite{Takeda/Yamaguchi/Watanabe:1972}} 
%%

%%
We begin by studying the (SrTiO$_3$)$_3$/(SrFeO$_3$)$_1$ superlattice and plot 
in Figure \ref{fig:3sto-1sfo-dos} our calculated local densities-of-states 
(DOS) for (a) cubic SrTiO$_3$, (b) cubic SrFeO$_3$, and (c) the superlattice. 
Our results for the bulk compounds are consistent with the literature:
In SrTiO$_3$ we obtain a $\sim$2~eV band gap between an O 2$p$ valence band and 
a Ti $3d$ conduction band. 
SrFeO$_3$ is nearly half-metallic with a calculated magnetic moment of 
3.8~$\mu_B$ per Fe atom, consistent with a high-spin $d^4$ 
electronic configuration. 
The Fermi level (dashed line at 0~eV) lies in a region of majority spin 
Fe $e_g^{\uparrow}$--O~2$p^{\uparrow}$ hybridized orbitals and in the 
O $2p^{\downarrow}$ valence band. 
The electronic structure of the superlattice is close to a superposition 
of its constituents, except for a reduction in the contribution of the O 
$2p$ states at $E_F$ making the superlattice half-metallic. 
While the use of the Hubbard $U$ method in DFT calculations is known to drive 
bands toward integer filling, the half-metallicity is robust for $U = 0$ to $8$~eV.
%and increasing $U$ also shifts some of the $e_g^{\uparrow}$ 
%character from $E_F$ to lower energy.
%%

%%
\begin{figure}[b]
\includegraphics[width=0.40\textwidth]{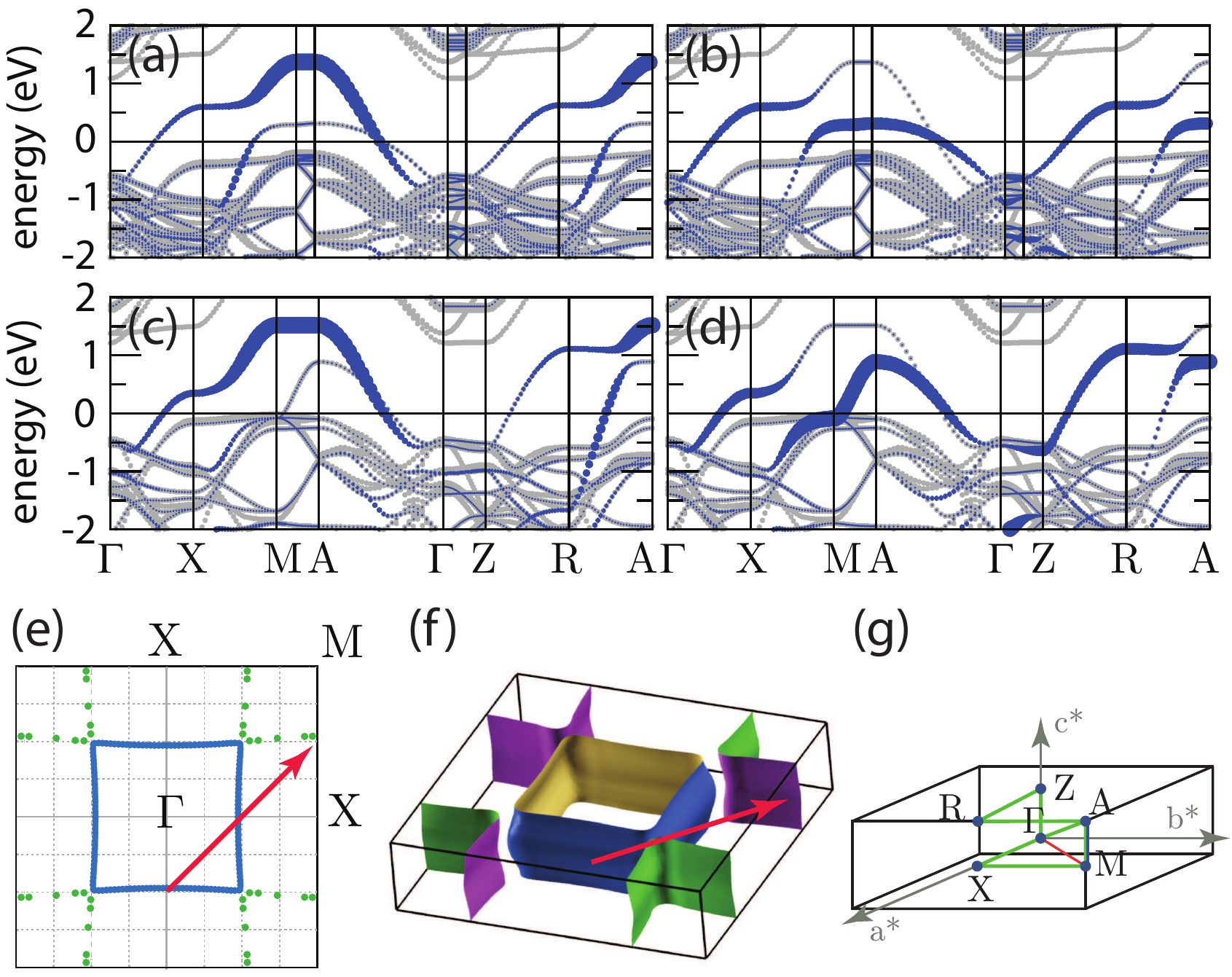}
\caption{\label{fig:comparison}(Color) 
Electronic band structure plots %along the high symmetry lines 
for the $n=3,m=1$ [panels (a) and (b)], and 
$n=1,m=1$ [panels (c) and (d)] superlattices.
Panels (a) and (c) show the majority spin ``fat-bands'' (blue) derived 
from  O 2$p_{x,y}$--Fe 3$d_{x^2-y^2}$ states, whereas  
(b) and (d) show the O 2$p_z$--Fe 3$d_{z^2}$ states.
The Fermi surface at $k_z=0$ (e) and in the full (f) BZ shown in (g)  
  consists of two bands %: 
%  an electron sheet (blue) centered at $\Gamma$ and a 
%  hole sheet (green) centered about M. 
  %
  connected by the ${\bf Q} = (1,1,0)\frac{\pi}{a}$  interband nesting vector (arrow).
  }
\end{figure}
We plot the band structure of the $P4/mmm$ superlattice in 
Figure \ref{fig:comparison}(a,b) 
using the ``fat-band'' method \cite{Jepsen/Andersen:1995}
in which the magnitude of the projection of each Bloch state onto a particular
set of atomic orbitals is represented by its line width: 
Panel (a) shows the equatorial O 2$p_{x,y}$--Fe 3$d_{x^2-y^2}$ states and (b) shows
the O 2$p_z$--Fe 3$d_{z^2}$ orbitals.
Consistent with the DOS, we find partial occupation of the majority 
$d_{x^2-y^2}$ and  $d_{z^2}$ Fe states at $E_F$ with the energy of the 
$d_{x^2-y^2}$ at $\Gamma$ 0.50~eV lower than that of the $d_{z^2}$ orbital.
%
%The tensile in-plane strain on SrFeO$_3$ reduces the $c$ axis length 
%compared to its bulk value; in fact we find that the apical 
%($z$-direction) Fe--O distance is reduced to 1.88~{\AA}.
%
The energy difference at  $\Gamma$ is explained from compression of 
the apical (1.88~{\AA}) over the in-plane  (1.93~{\AA}) Fe--O bond length from the 
imposed $\Delta_{\rm ECF}$ \cite{Hansmann/Held_et_al:2009}.
Interestingly, the apical Fe--O distance  is shorter than the 
value (1.90~{\AA}) that we obtain when we force {\it bulk} SrFeO$_3$ to have 
the SrTiO$_3$ in-plane lattice constant, suggestive of an additional spontaneous 
Jahn-Teller-like distortion (we examine this later). 
Despite the substrate elongation of the equatorial Fe--O bonds, 
the bands of $d_{x^2-y^2}$ character do not fully split from the $d_{z^2}$ 
bands to produce a single band crossing $E_F$; in fact, both the top and the 
bottom of the $e_g$ band are dominated by $d_{x^2-y^2}$ states.
The  $d_{z^2}$ band can likely be further destabilized to achieve an 
identical cuprate-like electronic structure %\cite{Pavarini/Andersen_et_al:2001} 
with additional strain \cite{Chaloupka/Khaliullin:2008} 
and/or chemical modifications \cite{Hansmann/Held_et_al:2009}.
We now investigate the character of the 2D bands 
and examine how the electronic structure at $E_F$ responds to 
changes in the dielectric thickness  
by comparing the full band structures of the (SrTiO$_3$)$_3$/(SrFeO$_3$)$_1$  
[Fig.~\ref{fig:comparison} (a,b)] and (SrTiO$_3$)$_1$/(SrFeO$_3$)$_1$ 
[(c,d)] superlattices. 
As expected, the bandwidth of the $d_{x^2-y^2}$ derived states along 
$\Gamma$--A is nearly the same in the two superlattices due to identical 
in-plane structural parameters.
The dispersion of the $d_{z^2}$ bands is markedly different, however, with the 
less confined $n=1,m=1$ superlattice having a larger bandwidth than the more 
confined $n=3,m=1$ along the  M--A lines.
Increasing dielectric thickness ($n=1\ldots4$) shows that the 
dispersive  $d_{x^2-y^2}$ bands saturate at $\sim$2~eV in width while the 
$d_{z^2}$ states are almost dispersionless, although never fully 
split.
In contrast, increasing the thickness of the 
SrFeO$_3$ layers in the superlattice produces several partially occupied 
$e_g$ bands crossing $E_F$, and the band structure (not shown) resembles a regular 
metal. 
We conclude that a cuprate-like band structure is unlikely in superlattices 
containing multiple ferrate layers.
The 2D confinement is discernible in the Fermi surface of  
the  $n=3,m=1$ superlattice shown in Fig.\ \ref{fig:comparison}(e,f).
The additional band crossing $E_F$ in the $k_z=0$ plane, not seen in 
the $n=1,m=1$ (c,d) superlattice nor bulk SrFeO$_3$ lattice matched to 
SrTiO$_3$, produces squared-cylindrical arrays around the M-point consistent 
with the four-fold symmetry of the lattice.
An unusual inward bowing along the $\Gamma$--X direction occurs with  
corrugations along $k_z$ similar to over-doped cuprates.
The band curvature indicates electron and hole sheets at $\Gamma$ and M, 
respectively,  
%due to the $d(x^2-y^2)$ [$d(z^2)$] band 
similar in character to the 
superconducting cuprates \cite{Cohen/Pickett/Krakauer:1988,Cohen_et_al:1990}.
Interestingly, the two Fermi sheets nearly intersect at half the 
$\Gamma$--M distance producing a nesting vector ${\bf Q} = (1,1,0)\frac{\pi}{a}$ 
connecting them.
Such a vector can result in a charge or spin density wave, or for strong 
coupling of the electronic system to the lattice, a symmetry lowering structural 
distortion. % which may gap the energy spectrum.  
These two sheets are largely insensitive to the calculation details: changing $U$ 
weakly affects the band crossings at $E_F$ since the  electronic structure 
is mainly modified by the confining SrTiO$_3$ layers. 
Spin-orbit interactions also do not alter the FS nesting. 
%%

\begin{figure}
\includegraphics[width=0.43\textwidth]{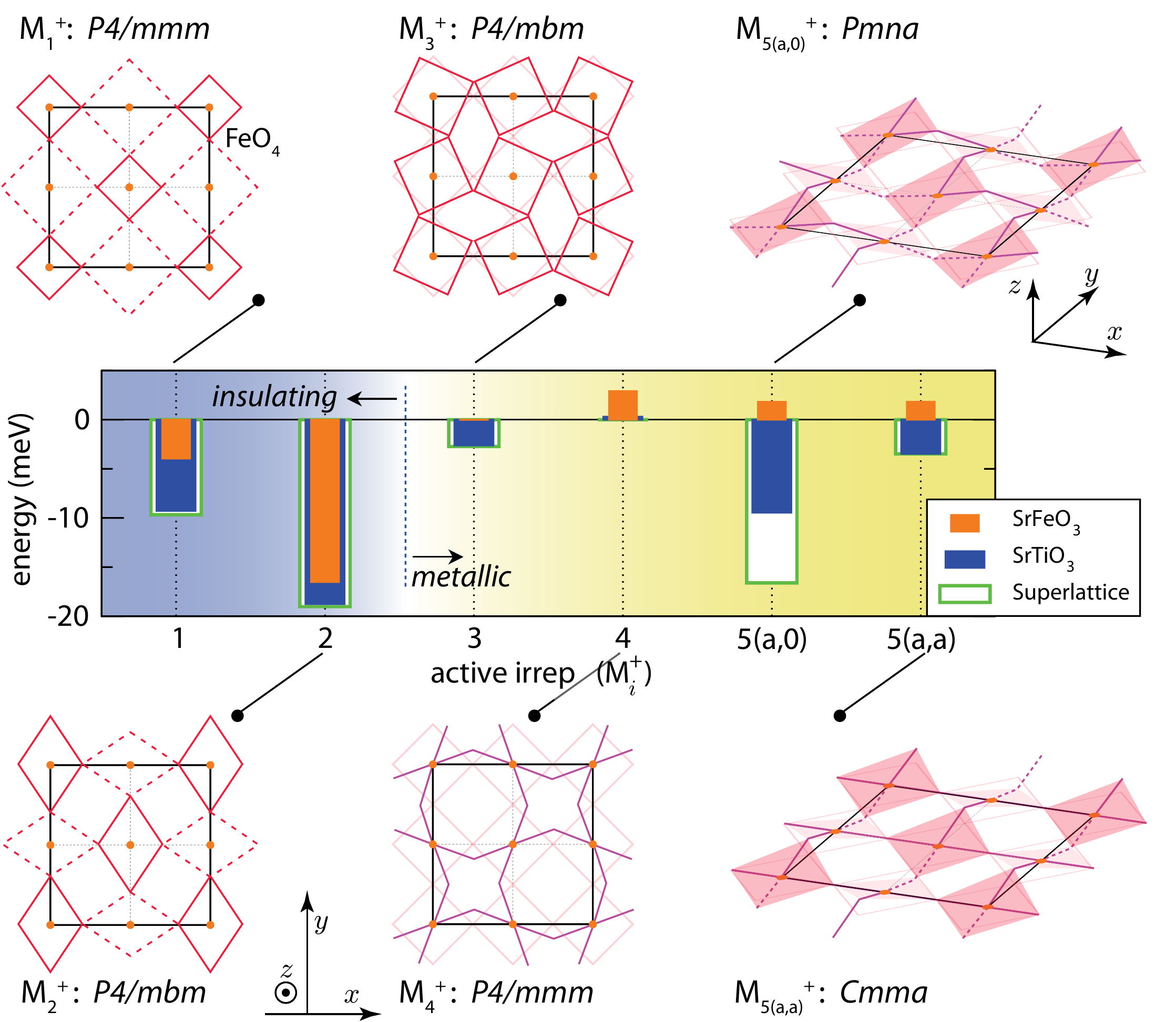}
\caption{\label{fig:irreps} (Color) %
  Electronic responses from SrFeO$_3$ (orange), SrTiO$_3$ (blue), 
  and the superlattice (green) to M$_i^+$ irreps of space group $P4/mmm$:
  M$_1^+$ breathing mode, 
  M$_2^+$ stretch mode, 
  M$_3^+$ rotation mode of the FeO$_6$ octahedra around the $z$-direction, 
  M$_4^+$ bending of the Fe--O bonds (purple), and two different tilt patterns 
  M$_5^+(a,0)$ and M$_5^+(a,a)$ in the $xy$-plane.
  }
\end{figure}

%%
%{\it Fermi surface topology and instabilities.} 
%%
We now investigate possible structural instabilities in the 
(SrTiO$_3$)$_3$/(SrFeO$_3$)$_1$ superlattice, and their influence 
on the FS degeneracy.
We identify likely M-point instabilities (consistent with the nesting vector) 
by choosing structural distortions that connect the high symmetry 
reference phase to low symmetry structures through atomic distortions that 
maintain a direct group-subgroup relation.
We choose these distortions  \cite{Campbell/Stokes:2006} to be irreducible 
representations (irreps) of space group $P4/mmm$ and consider only M$_i^+$, for 
$i=1\ldots5$. 
(We ignore  M$_i^-$ irreps with antisymmetric distortions 
under inversion, and follow the notation of 
Miller \& Love throughout \cite{Miller/Love:1967}.)
Fully relaxed {\it ab initio} structures are then 
reduced into combinations of these irreps by performing a symmetry mode 
analysis \cite{Aroyo/Perez-Mato_et_al:2006} %
%,Orobengoa/Perez-Mato_et_al:2009
that makes accessible the local atomic displacements.
We keep FM spin order fixed as above to isolate the electron-lattice coupling.
The active irreps we examine (Fig.\ \ref{fig:irreps}) affect either the 
equatorial Fe--O bond lengths (crystal field) or O--Fe--O bond angles 
($dp\sigma$ bandwidth).
Irreps M$_1^+$ and M$_2^+$ are planar breathing and stretch modes of 
the FeO$_4$ plaquettes, respectively, and are anticipated to 
strongly modulate the $e_g^\uparrow$--O~2$p$ hybridization at $E_F$.
The M$_1^+$ mode produces two unique FeO$_4$ plaquettes arranged 
in a 2D checkerboard manner and gives rise to 
charge disproportionation in bulk isoelectronic 
CaFeO$_3$ \cite{Matsuno/Fujimori_et_al:2002}.
In contrast the M$_2^+$ irrep, or Jahn-Teller mode, creates two long and 
two short equatorial Fe--O bonds and favors orbital polarization.
% as in many $d^4$ manganate oxides.
%
The remaining irreps are collective distortions of the octahedral units that 
produce the common tilt and rotation patterns found in perovskite oxides.
These modes affect the $dp\sigma$ bandwidth through deviations 
in the Fe--O--Fe bond angle away from ideal 180$^\circ$:
M$_3^+$ consists of rotations of the in-plane oxygen atoms about the $z$-axis;  
M$_4^+$ produces a bending distortion of the O--Fe--O bond; and 
the 2D irrep M$_5^+$ produces two possible tilt patterns of the planar oxygen atoms: 
M$_5^+$(a,0)  giving a pattern with space group symmetry $Pmna$ or 
M$_5^+$(a,a) with $Cmma$.
We begin by ``freezing'' each irrep as a function of 
mode amplitude into the equatorial oxygen atoms coordinating Fe 
% 
%[We increase the size of superlattice to be commensurate with the 
%distortions  (see Supp. Materials).]
(see Supp. Materials).
At the same time, we keep the remaining atoms in the superlattice 
clamped to the reference $P4/mmm$ positions to isolate 
the response of the ferrate layer---displacements of other atoms in 
the superlattice are obtained later through full structural optimization.
The energy change due to the optimal mode amplitude for each irrep  %\footnote{
%
%A distortion of 0.05~{\AA} is chosen for irreps which increase the energy 
%of the superlattice.}
%
frozen in the ferrate layer is shown in Fig.\ \ref{fig:irreps} with 
respect to the high symmetry $P4/mmm$ superlattice (energies per 
five-atom perovskite cell).
Irreps M$_1^+$, M$_2^+$ and M$_3^+$ lower the energy of the 
{\it superlattice}. In contrast they increase the energy of 
{\it bulk} SrFeO$_3$ lattice matched to SrTiO$_3$, indicating 
that the superlattice geometry makes the  
$e_g$ orbital degeneracy more susceptible to structural distortions.
The remaining irreps which require collective cooperation of all 
oxygen octahedra increase the superlattice energy when the dielectric is 
clamped to the reference configuration.
Next, we fully relax the SrTiO$_3$ with the ferrate atoms fixed to the 
distortions above % within the symmetry constraints imposed by each irrep 
to isolate the effect of the dielectric response.
%Using the relaxed structures we decompose the structure in linear combinations 
%of the irreps and plot the energy difference attributed to the dielectric response.
%
%Some variation in the irrep amplitude arises from condensation of a 
%$\Gamma_1^+$  mode. (Complete details of the distortion vectors are given 
%elsewhere.\cite{epaps:2009})
%
Now every irrep except M$_4^+$ is found to be energy lowering: 
the large energy gain now observed for the M$_5^+$ modes 
suggests that a 3D tilt pattern is preferred---this result is likely driven by 
the fact that most of the superlattice is SrTiO$_3$ which itself has a 
large M$_5^+$-like mode instability. 
% itself adopts a  tetragonal %$I4/mcm$ 
%crystal structure with deviations in the Ti--O--Ti angles. 
% driven by softening of a R$_4^+$ irrep of the cubic $Pm\bar{3}m$ perovskite cell.
%%

\begin{figure}
\includegraphics[width=0.44\textwidth]{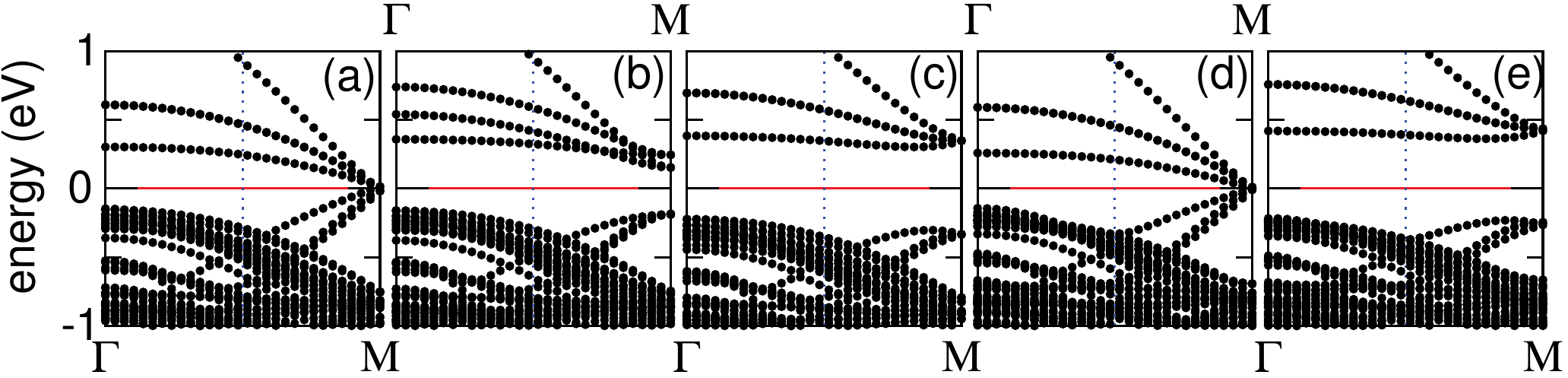}
\caption{\label{fig:gm-summary}  Band structure for (a) the $P4/mmm$  
$(n=3,m=1)$ superlattice as in Fig.\ \ref{fig:comparison}(a) with the 
lattice parameter doubled in-plane to allow comparison with 
panels (b)-(d) which include additional structural distortions:
(b) M$_1^+$, (c) M$_2^+$, and (d) M$_5^+(a,0)$; and the ground state $P2_1/c$ 
structure (e).}
\end{figure}

%%
We now relax both constituents under the symmetry constraints imposed by the irreps 
to find the lowest energy atomic configurations.
The largest energy-lowering distortions are found to be the Jahn-Teller and 
tilt modes. %, with the former more favorable than the breathing distortion which 
%also directly modifies the $e_g$ states.
%
%This can be understood as a volume conservation effect, where the Jahn-Teller mode 
%nearly conserves the volume of the octahedra (less than 0.40\% different 
%from the reference structure compared to 8.1\% different for the breathing distortion) 
%and thus preserves the registry of the superlattice.
%
%The breathing mode (M$_1^+$) produces two FeO$_6$ octahedra which 
%differ in volume by 8.1\% due to the different equatorial bond 
%lengths of 1.96 and 1.90~{\AA}.
%
%In contrast, the stretch mode produces planar Fe--O bonds of 2.01 and 1.86~{\AA}, 
%with a octahedral volume less than 0.40\% different from the reference structure.
%
Both irreps M$_1^+$ and M$_2^+$, open energy gaps at 
the Fermi level in the superlattice [Fig.\ \ref{fig:gm-summary}(b,c)],
while identical distortions in bulk SrFeO$_3$ maintain its metallicity.
The tilt modes M$_5^+$(a,0) and M$_5^+$(a,a) are more unstable than the 
rotation M$_3^+$ irrep with M$_5^+$(a,0) almost as energetically favorable as 
the Jahn-Teller distortion.
%
%In this case, the the Fe--O--Fe in-plane bond angle is reduced by 5.7$^\circ$ and 
%the out-of-plane Fe--O--Ti bond angle is also reduced by 8.6$^\circ$. 
%
Yet, due to the weaker influence of the bond angle on the $dp\sigma$ bandwidth, 
none of the bond angle distortions gap the FS.
There is a 0.24~eV bandwidth reduction, however, for irrep M$_5^+$ shown in 
Fig.\ \ref{fig:gm-summary}(d). 
Similarly, the M$_1^+$ mode  also gaps the  Fermi surface---although not an 
energy lowering distortion---in doped La$_2$CuO$_4$,  whereas 
the octahedra tilt and rotation modes, which are found in the cuprates, show 
minor effects \cite{Mattheiss:1987}.
Finally, we calculate the fully optimized structure starting from a 
combination of the M$_2^+$ and M$_5^+$ irreps.
Our ground state structure is 33.5~meV lower in 
energy (per 5-atom unit cell) than the reference superlattice.
It contains FeO$_4$ plaquettes with bond lengths of 1.86 and 2.01~{\AA} 
due to the Jahn-Telller mode thats makes the superlattice insulating 
[0.58~eV gap, Fig.\ \ref{fig:gm-summary}(e)].
The octahedral instabilities largely condense in 
the SrTiO$_3$ layers with reduced interaction in the ferrate plane: 
The in-plane Fe--O--Fe bond angle is reduced by 6.89$^\circ$ while a 
10$^\circ$ reduction occurs across the Fe--O--Ti angle.  
By symmetry decomposing the structure as a combination of the $P4/mmm$ irreps, $0.051\Gamma_1^+ + 0.048\Gamma_5^+ + 0.21$M$_2^+ + 0.13$M$_3^+ + 0.97$M$_5^+$,  
%(see Supp.\ Materials)  
%
%%By considering group-subgroup relations,\cite{Carpenter/Howard:2009} 
%This distorted structure evolves from condensation of 
we find the primary order parameters  M$_2^+$ and M$_5^+$ drive the  
%a second order phase transition from the metastable 
$P4/mmm~(a^0,a^0,c^0) \rightarrow P2_1/c~(a^+,b^-,b^-)$ metal-insulator 
transition, with the associated change in octahedral tilt patterns. 
%
%After this distortion, the $\Gamma_5^+$ and M$_3^+$ components are allowed 
%to be non-zero without further symmetry reduction thus making them secondary order parameters.
%Interestingly, the orbital-ordered Jahn-Teller manganate and vanadate 
%compounds adopt the same ground state crystal structure.
%
%M$_3^+$ however is restricted to the TiO$_4$ layers producing a rotation of 0.88$^\circ$.
%
Since none of these distortions are favored in bulk SrFeO$_3$, 
we attribute the  electron-lattice coupling enhancement to the 
confinement effects imposed by the superlattice geometry. 
%%

%{\it Conclusion.}
%
We have demonstrated that in a superlattice geometry, confinement from 
dielectric spacer layers combines with a strain-induced epitaxial crystal 
field ($\Delta_{\rm ECF}$) to modify %the otherwise 
%degenerate states near the Fermi energy in $d^4$ SrFeO$_3$.
%
the energies and dispersions of apical $d_{z^2}$ and in-plane Fe $d_{x^2-y^2}$ 
orbitals.
When we simulate the high symmetry cubic structures of room temperature 
SrTiO$_3$ and SrFeO$_3$ partial planar electron localization yields a 
2D Fermi surface that resembles  the superconducting cuprates.
%; this structure is preserved even in superlattices 
%terminated with a monolayer of SrO.
%
Low-energy M-point instabilities compete with nesting on the Fermi surface 
to make the superlattices proximal to multiple competing phases.
The lattice instabilities are enhanced by the dielectric 
SrTiO$_3$ layers, therefore inert spacer layers without 
polarizable ions might be more favorable for superconductivity.
We hope that our finding of competing structural ground 
states with distinct electronic properties motivates experimental 
investigation of SrFeO$_3$/SrTiO$_3$ heterostructures using 
external electric and magnetic fields:
Such probes could tune between the itinerant and localized electronic 
states in analogy with the parent Mott insulating cuprates.
%%

%{\it Acknowledgments.} 
We gratefully acknowledge support from NDSEG (JMR), the 
NSF under grant no.\ DMR 0940420 (NAS), and discussions with  
C.\ Adamo and K.\ Delaney.
%, O.K.\ Andersen, and W.\ Pickett.
%
%Portions of this work made use of the SGI Altix {\sc cobalt} system 
%at the National Center for Supercomputing Applications under grant 
%no.\ TG-DMR-050002S, and the CNSI Computer Facilities at UC Santa Barbara
%under NSF award no.\ CHE-0321368.

%%
% Bibtex file here
%\bibliographystyle{apsrev_etal1}
%\bibliography{sfo}

% Phew...EOF
\end{document}